\documentclass[aps,prl,reprint,floatfix,groupedaddress,superscriptaddress,showpacs,longbibliography]{revtex4-2}

\usepackage{lipsum}
\usepackage{graphicx}
\usepackage{natbib}
\usepackage[utf8]{inputenc}
\usepackage{tikz}
\usepackage{amsmath}
\usepackage{empheq}
\usepackage{geometry}
\usepackage{amssymb}
\usepackage{xcolor}
\usepackage{soul}

\newcommand{\srni}{Sr$_\text{3}$NiRhO$_\text{6{}{}{}}$}

\begin{document}
	
	\title{Magnetic ground state and excitations in mixed 3$d$-4$d$ quasi-1D spin-chain oxide \srni}
	
	\author{A.~Jain}
	\email{ajain@barc.gov.in}
	\affiliation{Solid State Physics Division, Bhabha Atomic Research Centre, Mumbai 400085, India}
	\affiliation{Homi Bhabha National Institute,   Anushaktinagar, Mumbai 400 094, India}

	\author{D. T. Adroja}
	\email{devashibhai.adroja@stfc.ac.uk}
	\affiliation{ISIS Facility, STFC, Rutherford Appleton Laboratory, Chilton, Oxfordshire OX11 0QX, UK}
	\affiliation{Highly Correlated Matter Research Group, Physics Department, University of Johannesburg, PO Box 524, Auckland Park 2006, South Africa}

	\author{S. Rayaprol}
	\affiliation{UGC-DAE CSR, Mumbai Center,  CFB, BARC, Trombay, Mumbai 400085, India}

	\author{A. D. Hillier}
	\affiliation{ISIS Facility, STFC, Rutherford Appleton Laboratory, Chilton, Oxfordshire OX11 0QX, UK}

	\author{W. Kockelmann}
	\affiliation{ISIS Facility, STFC, Rutherford Appleton Laboratory, Chilton, Oxfordshire OX11 0QX, UK}

	\author{S. M. Yusuf}
	\affiliation{Solid State Physics Division, Bhabha Atomic Research Centre, Mumbai 400085, India}
	\affiliation{Homi Bhabha National Institute,   Anushaktinagar, Mumbai 400 094, India}

	\author{E. V. Sampathkumaran}
	\affiliation{Department of Physics, Indian Institute of Technology Roorkee, Roorkee 247667, India}
	\affiliation{Tata Institute of Fundamental Research, Homi Bhabha Road, Colaba, Mumbai 400005, India}

	\date{\textrm{\today}}
	
	\begin{abstract}
		
		Entanglement of spin and orbital degrees of freedom, via relativistic spin-orbit coupling, in 4$d$ transition metal oxides can give rise to a variety of novel quantum phases.  A previous study of mixed 3$d$-4$d$ quasi-1D spin-chain oxide \srni\ using the magnetization measurements by Mohapatra \textit{et al.} [Phys. Rev. B 75, 214422 (2007)] revealed a partially disordered antiferromagnetic (PDA) structure below 50 K ~\cite{Mohapatra2007}. We here report the magnetic ground state and spin-wave excitations in  \srni\,  using muon spin rotation and relaxation ($\mu$SR), and neutron (elastic and inelastic) scattering techniques. Our neutron diffraction study reveals that in the magnetic structure of \srni\,  Rh$^{4+}$ and Ni$^{2+}$ spins are aligned ferromagnetically in a spin-chain, with moments along the crystallographic $c$-axis. However, spin-chains are coupled antiferromanetically in the $ab$-plane. 
		$\mu$SR  reveals the presence of oscillations in the asymmetry-time spectra below 50 K, supporting the long-range magnetically ordered ground state. Our inelastic neutron scattering study reveals gapped quasi-1D magnetic excitations with a large ratio of gap to exchange interaction. The observed spin-wave spectrum could be well fitted with a ferromagnetic isotropic exchange model (with $J = 3.7 $ meV) and single ion anisotropy ($D=10$ meV) on the Ni$^{2+}$ site. The magnetic excitations survive up to 85\,K, well above the magnetic ordering temperature of $\sim 50$\,K, also indicating a quasi-1D nature of the magnetic interactions in \srni. 
	\end{abstract}
	
	\maketitle
	\setcounter{secnumdepth}{1}
	\section{Introduction}
\begin{figure}[!htb]
\centering
\includegraphics[width = \linewidth]{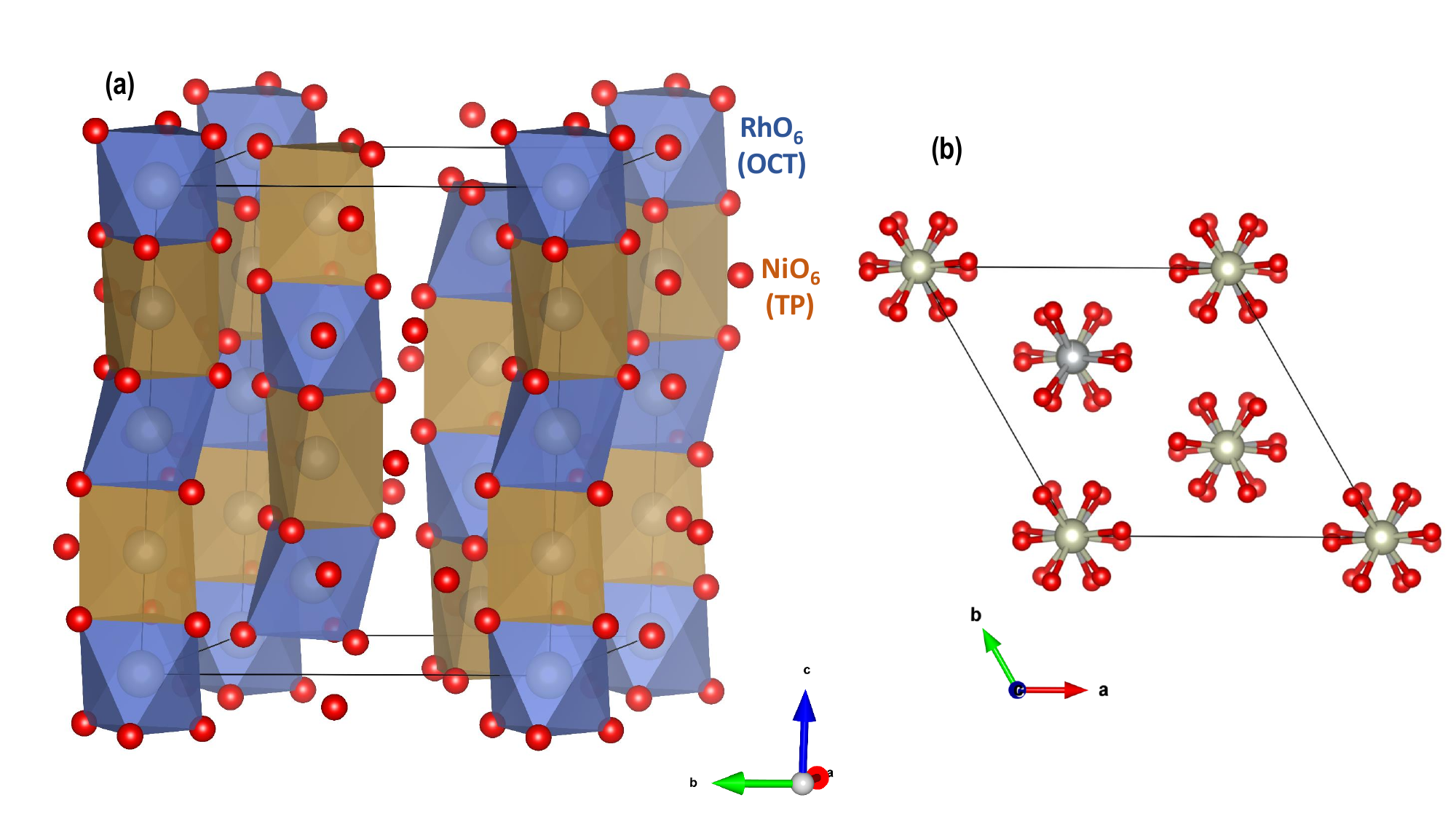}
\caption{(Color online)  Crystal structure of \srni: A perspective view showing 1D spin-chains made up of alternating face-sharing {NiO$_{6}$ trigonal prisms  and RhO$_{6}$  octahedra}  (left). Right side shows triangular lattice arrangement of spin-chain in $ab$-plane.  Red balls show O atoms. For clarity strontium atoms are not shown.}
\label{fig1:crystst}
\end{figure}

	Quasi-one-dimensional (1D) quantum spin-chain compounds have attracted a lot of attention in the last two decades due to a variety of interesting quantum phenomenon revealed by them. Among these, spin-chain compounds with general formula $A_{3}MX$O$_{6}$ ($A$ here denote Sr, Ca, etc. and $M$, $X$  denote transition metals) have attracted much attention in recent years ~\cite{Mohapatra2007,Nguyen1995, AASLAND1997187,Niazi2001, Sampathkumaran2002, Niazi2002b, Niazi2002, Niazi2002b, Loewenhaupt2003, Flahaut2003, Flahaut2003, Rayaprol200379, Sampathkumaran2004, Sampathkumaran2004a, Sampathkumaran2004, Hardy064424,Takubo2005, Wu2005, JainPhysRevB.74.174419, 	Sampathkumaran2007,	Takubo2005, Agrestini2008, PhysRevLett047601,	Chapon2009, PhysRevLett187202, Sarkar2010,	Hillier2011, Agrestini197204, JainPhysRevB.83.184425, Moyoshi034701, Mikhailova2012, Jain2013, Basu2013, Yin2013a, Yusuf_2013, JainPhysRevB.87.094411, Ou2014,	Lefrancois2014, Agrestini2014, 	Basu2014,  	Rayaprol200379, Paddison014411, Toth2016,  McClarty064703, Jain184406}, due to their peculiar properties such as time dependent magnetic order~\cite{Agrestini197204, Jain184406},  steps in the dc magnetization~\cite{Hardy064424}, order by disorder~\cite{PhysRevLett187202},  and multiferroicity~\cite{PhysRevLett047601}. The crystal structures of these compounds contain spin-chains, made up of alternating face-sharing  {$X$O$_{6}$ octahedra (OCT) and  $M$O$_{6}$ trigonal prisms (TP)}, running along the crystallographic $c$-axis (Fig.~\ref{fig1:crystst}) ~\cite{Nguyen1995}. These chains are arranged on a triangular lattice in the $ab$-plane. In most of the compounds, where 3$d$ transition metal ions occupy both OCT and TP sites, intrachain magnetic exchange interaction is ferromagnetic (FM)~\cite{Jain2013,Paddison014411}. This combined with an antiferromagnetic (AFM) interchain interaction and triangular lattice arrangement of spin-chains  gives rise to  geometrically frustrated magnetism~\cite{Jain2013,Agrestini2014,Agrestini2008,Sampathkumaran2004a,Sampathkumaran2007, Paddison014411}. Among these compounds, the cobalt base system  Ca$_3$Co$_2$O$_6$ has been extensively investigated in last two decades~\cite{Jain2013,Agrestini2014,Agrestini2008,Sampathkumaran2002,Sampathkumaran2004,Sampathkumaran2004a,Yusuf_2013,Takubo2005,Sampathkumaran2007, Rayaprol200379, JainPhysRevB.83.184425,Jain184406, Paddison014411}. Due to spin-orbit coupling ($\sim 70\,$meV) cobalt ions in   Ca$_3$Co$_2$O$_6$  show Ising-like behaviour, with spin aligned ferromagnetically along the crystallographic $c$-axis.
	
	Initial interest in the compounds, where 3$d$ transition metal ions occupy both OCT and TP sites, was motivated because  correlation effects are more prominent in 3$d$ transition metal than for the 4$d$ and 5$d$ transition metal ions. 
	However, discovery of a Mott insulating state in Sr$_2$IrO$_4$ in the late 2000s~\cite{Bjkim2009} and the role of spin-orbit coupling (SOC) in stabilizing the Mott state awoke an interest in the  compounds where spin-chains were made up of 3$d$ and 5$d$ spin-chain oxides. 
Subsequently, many Sr-based compounds of the $A_{3}MXO_{6}$ family, namely Sr$_3$$M$IrO$_6$ with $M$=Co, Cu, Ni, and Zn, were investigated~\cite{Niazi2001, Niazi2002b, Niazi2002, Niazi2002b, Flahaut2003, Mikhailova2012, Sarkar2010, Ou2014,McClarty064703, Lefrancois2014, Toth2016}. For Sr$_3$CuIrO$_6$~\cite{Yin2013a},  a large gap in the magnetic excitation spectrum was  observed which was ascribed to an unusual exchange anisotropy generating mechanism, namely, strong FM anisotropy arising from AFM super-exchange interaction, driven by the alternating weak and strong spin-orbit coupling on the 3$d$ Cu and  5$d$ Ir magnetic ions, respectively. Non-collinear AFM ordering has been observed in Sr$_3$ZnIrO$_6$ and Sr$_3$ZnRhO$_6$ which has been ascribed to anisotropic intra-chain and inter-chain exchange interactions~\cite{Hillier2011,McClarty064703}, arising due to a strong  spin-orbit coupling. In Sr$_3$NiIrO$_6$, anisotropic symmetric interaction from strong spin-orbit coupling was theoretically proposed to  result in a strong magnetic anisotropy and AFM intrachain interactions~\cite{Birol134432}.  Therefore, in search of  unusual magnetic behavior expected in the mixed 3$d$ and 4$d$ transition-metal compounds via exchange pathways that are absent in the compound with only  3$d$ transition-metal ions, some of us reported interesting magnetic anomalies with partially disordered antiferromagnetism  in \srni, based on bulk magnetization techniques~\cite{Mohapatra2007}.  This kind of magnetism has garnered significant interest in recent years, as the magnetic ions of a given element occupy crystallographically equivalent positions, but only some of these ions exhibit magnetic ordering while others remain magnetically disordered due to geometrical frustration.  Unlike spin liquids, where disorder is uniform and intrinsic, partially disordered magnets feature order and disorder coexisting in the same magnetic phase, resulting in unique physical properties and potential applications.

We here present the results of an extensive investigation of the magnetism of \srni \,using  muon spin rotation and relaxation ($\mu$SR), elastic and inelastic neutron scattering techniques,  to gain a deeper microscopic insight into the nature of geometrically frustrated magnetism in this compound.
The present study firmly establishes that this compound orders in  a partially disordered antiferromagnetic (PDA) state below $\sim$ 50\,K ($T_1$). The PDA state undergoes a transition to a frozen PDA (F-PDA) below $\sim20$\,K ($T_2$). Further, our  inelastic neutron scattering study confirms the 1D nature {of the magnetism} of this compound.

\section{Experimental Details}
Polycrystaline samples were synthesized by the standard solid state reaction method starting
with high purity materials SrCO$_3$ (99.99$\%$), NiO (99.99$\%$), and RhO$_2$ (99.9$\%$) in powder form, as discussed in ref.~\cite{Mohapatra2007}.  The magnetization measurements (dc and ac) were carried out using a Quantum Design SQUID VSM to confirm the quality of the specimen magnetically vis-a-vis the ones reported in Ref.~\cite{Mohapatra2007} [See Appendices~\refeq{sec:appen} and ~\refeq{sec:Acs}]. The neutron diffraction experiments at 297~K were carried out using the five linear position sensitive detector (PSD) based powder diffractometer ($\lambda = 1.249 {\AA}$) at Dhruva research reactor, Trombay. Neutron diffraction experiments in the temperature range 5-100~K were performed using the time-of-flight GEM diffractometer at the ISIS facility, UK. 
The diffraction data were analyzed by the Rietveld  method, using the FULLPROF program~\cite{Rodriguez-carvajal1993}. The inelastic neutron scattering (INS) measurements over 5-300 K  were performed using the time-of-flight chopper spectrometer HET at the ISIS Neutron and Muon Facility, UK. Muon Spin Rotation and Relaxation ($\mu$SR) experiments were carried out using the MuSR spectrometer at the ISIS Neutron and Muon Facility, UK.

\section{Results}
\subsection{A. Crystal Structure: Neutron Diffraction}

\begin{figure}[h]
\centering
\includegraphics[width = 0.95\linewidth]{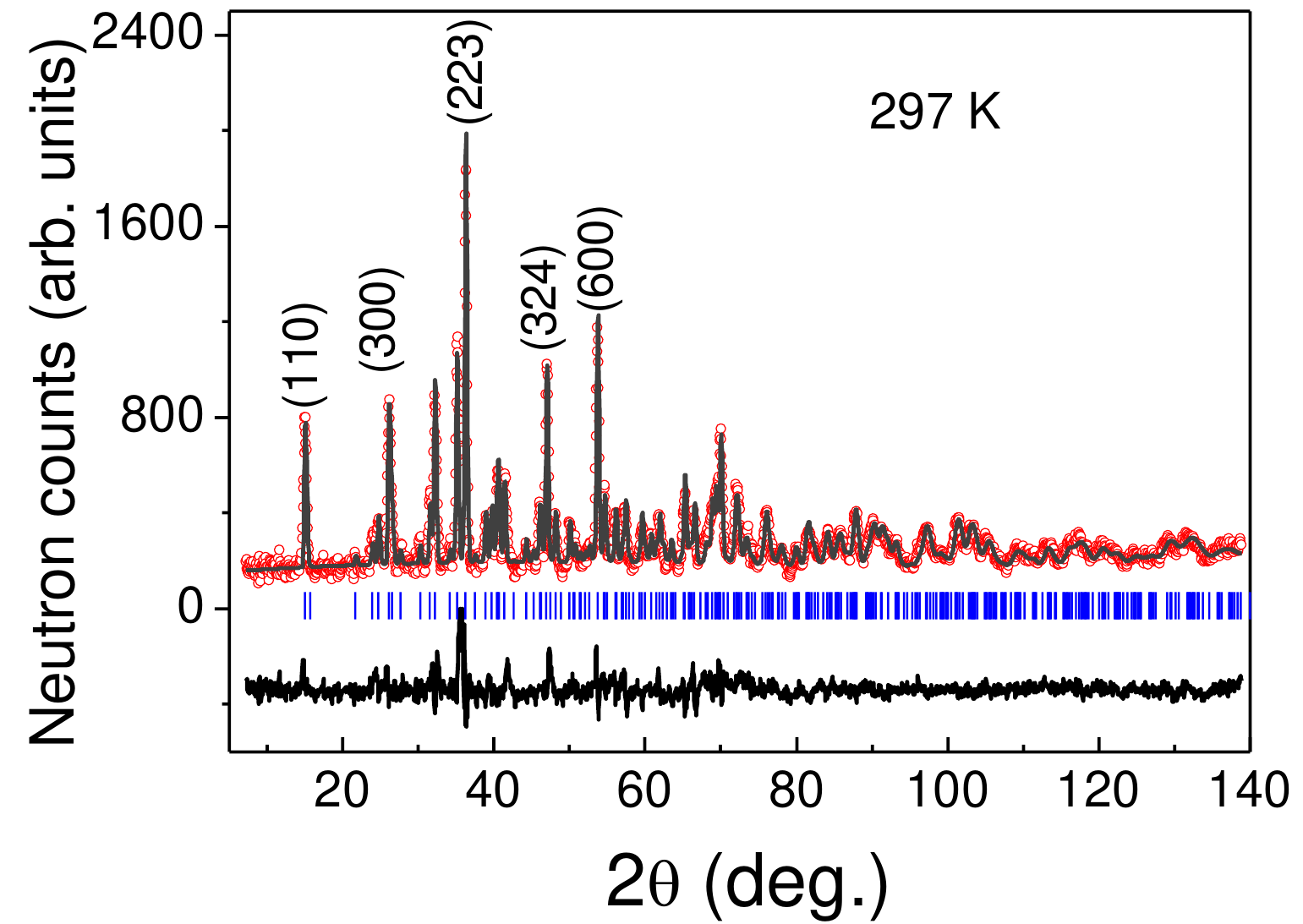}
\caption{(Color online) Observed (data points) and calculated (solid line through data the data points) neutron diffraction patterns for \srni{} at 297~K using a wavelength of 1.249~${\AA}$. The black solid line at the bottom shows the difference between the observed and the calculated patterns. Vertical lines show the positions of the nuclear Bragg peaks. The ($hkl$) values corresponding to a few strong Bragg peaks are also listed.}
\label{Fig2:Dhruva}
\end{figure}

Figure~\ref{Fig2:Dhruva} depicts the Rietveld refined neutron diffraction pattern at 297 K for \srni{} measured at Dhruva research reactor using a wavelength of 1.249~\AA.  The refined values of the lattice constants $a = $9.585(1), and $c =$ 11.041(1)~{\AA} (space group $R  \overline 3 c$) are found to be in good agreement with those reported in the literature~\cite{Stitzer2002, Stitzer2002, Vajenine1996, Mohapatra2007}. There are six formula units ($Z= 6$) of \srni{}   in a crystallographic unit cell. The refinement also confirms that Ni$^{2+}$ and Rh$^{4+}$ ions are located at the {TP  and OCT} sites, respectively. {We have also performed a detailed analysis of
neutron diffraction pattern at 100~K, collected using the time-of-flight GEM diffractometer at the ISIS facility, UK, covering a wide $Q$-range (Fig.~\ref{Fig13:GEM}, Appendix~C). The unit cell parameters and other structural parameters
obtained from the refinement of the neutron diffraction patterns
at 100~K are given in Table~\refeq{sttable}.}


\begin{table*}[!t] \begin{center}
	{
		\caption{\label{sttable}Structural parameters (atomic positions, thermal parameters, obtained from the refinement of the neutron diffraction patterns at 100~K for \srni{}. The lattice constants are \emph{a} = 9.569(1) {\AA}, \emph{c} = 11.025(2) {\AA}.  All crystallographic sites are fully occupied.}}
		\vspace{0.2 cm}
		
		\begin{tabular}{p{1.5cm}cccccc}
			\hline
			\hline
			Atom  & \multicolumn{1}{c}{Site} & $\emph{x}$     & \emph{y}     & \emph{z} & \emph{B}$_{iso}(${\AA}$^{2}$) \\
			\\ [-1.5ex]
			\hline
			Sr & {18$\emph{e}$} & 0.3661(5) & 0     & 0.25  & 0.29(2) \\
			Rh &  {6$\emph{b}$} & 0     & 0     & 0     & 0.28 (3) \\
			Ni   & \multicolumn{1}{c}{6$\emph{a}$} & 0     & 0     & 0.25  & 0.23(3) \\
			O     & \multicolumn{1}{c}{36\emph{f}} & 0.1720(1)  & 0.0209(3)  & 0.1129(2) & 0.33(3)  \\   
			\hline
			\hline
		\end{tabular} 
	\end{center}
\end{table*}

\subsection{B. Magnetic Structure: Neutron Diffraction}
\begin{figure}[htb]
	\centering
	\includegraphics[width = \linewidth]{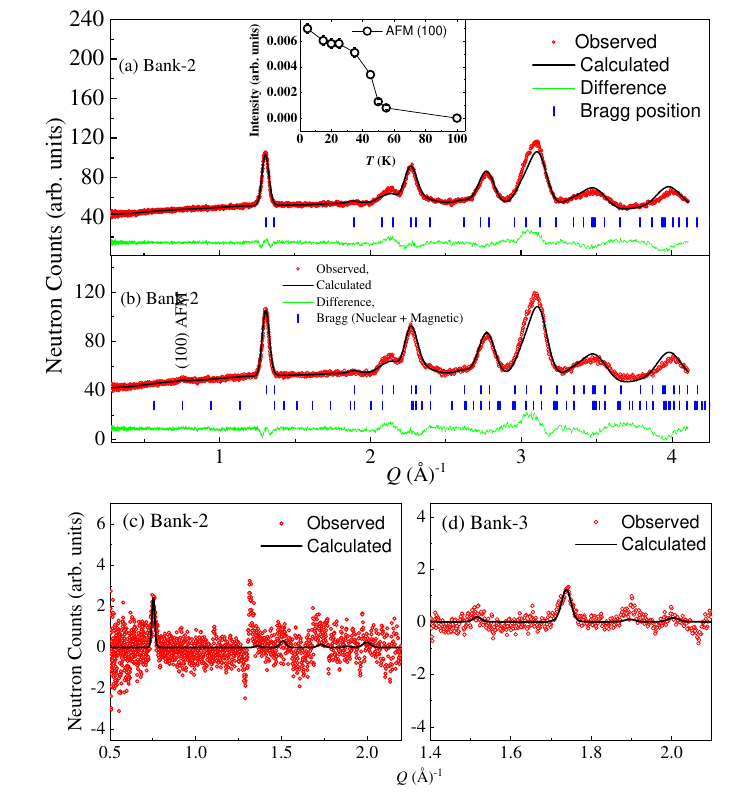}
	\caption{(Color online) Rietveld refined neutron diffraction patterns at (a) 100 and (b) 5~K from the detector bank-2 of the GEM diffractometer. Upper (lower)  vertical lines show the positions of the nuclear (magnetic) Bragg peaks. Observed (open circles) and calculated (solid line) neutron diffraction patterns.  Solid green line at the bottom shows the difference between the observed and the calculated patterns.  The inset in (a) shows the temperature dependent intensity of the magnetic Bragg peak. Observed  (open circles) neutron powder diffraction  pattern at 5~K after subtraction  of the nuclear pattern at 100~K and calculated (solid lines) pattern considering only magnetic phase with zero background from the  (c) detector bank-2 and  (d) detector bank-3  over low values of $Q$ from the GEM TOF diffractometer.
    The GEM detector banks 2 and 3 are positioned at scattering angles of 17.980° and 34.960°, respectively. Due to their distinct scattering angles, each detector bank has a different scaling factor. These scaling factors were estimated by fitting the data collected at 100 K.}
	\label{fig:diff}
\end{figure}

To investigate the magnetic ground state of \srni{}, we have performed neutron diffraction experiments over 5-100~K. Fig.~\ref{fig:diff}(a-b) shows Rietveld refined neutron diffraction patterns obtained from the GEM instrument at 5 and 100~K.  The neutron diffraction patterns at 100~K  can be fitted by considering only the nuclear phase.  The nuclear peaks as inferred from 100~K data are shown in (Fig.~\ref{fig:diff}(a)) to compare with the pattern shown in (Fig.~\ref{fig:diff}(b)).	As shown in  Fig.~\ref{fig:diff}(b), an additional peak in the neutron diffraction pattern is clearly visible at $Q\sim$ 0.76~{\AA}$^{-1}$ at 5~K compared with the 100~K data. The temperature dependence of the integrated intensity of the additional Bragg peak is shown in the inset of Fig.~\ref{fig:diff}(a). 
{The difference pattern between 5 and 100~K [Figs.~\ref{fig:diff}(c-d)], obtained from the detector bank-2 and bank-3, highlights distinct magnetic Bragg peaks, indicating long-range magnetic ordering. The propagation vector for the observed magnetic peaks is $k$ =(0, 0, 1) in the hexagonal setting of the space group $R\bar3c$. In the primitive rhombohedral cell setting, Kovalev’s notation,  $k$=(1/3, 1/3, 1/3)}.
Analysis of magnetic structure has been carried out using the standard irreducible representation theory with help of software BASIREPS~\cite{Rodriguez-carvajal1993}. In \srni, {Rh$^{4+}$ and Ni$^{2+}$}  ions are located at the 6\emph{b} (0, 0, 0) and 6\emph{a} (0, 0, 1/4) sites, respectively.    The magnetic reducible representations for these sites can be decomposed as direct sum of irreducible representations as
\begin{equation}\label{NEW6b}
	\Gamma(6\emph{b})= 1 \Gamma_{1}^{(1)} + 1 \Gamma_{2}^{(1)} + 2 \Gamma_{3}^{(2)}
\end{equation}
and
\begin{equation}\label{NEW6a}
	\Gamma(6\emph{a})= 1 \Gamma_{1}^{(1)} + 1 \Gamma_{2}^{(1)} + 2 \Gamma_{3}^{(2)}.
\end{equation}

The representations $\Gamma$$_{1}$ and $\Gamma$$_{2}$ are one-dimensional, while representation $\Gamma$$_{3}$ is two dimensional.
In the formalism of propagation vectors,  the magnetic moment of the atom ``\emph{j}" in the asymmetric magnetic unit cell with the origin at the lattice point \emph{R}$_{L}$, can be written as the Fourier series of the form

\begin{equation}\label{eq1parent}
	\textbf{\emph{m}}_{L,j} =  \sum_{\textbf{k}} \textbf{S}_{j}^{k}\mbox{ } exp ({- 2\pi \emph{i} \textbf{\emph{k}}. \textbf{\emph{R}}_{L}}).	
\end{equation}

\begin{figure}[!htb]
	\centering
	\includegraphics[width = 1.01\linewidth]{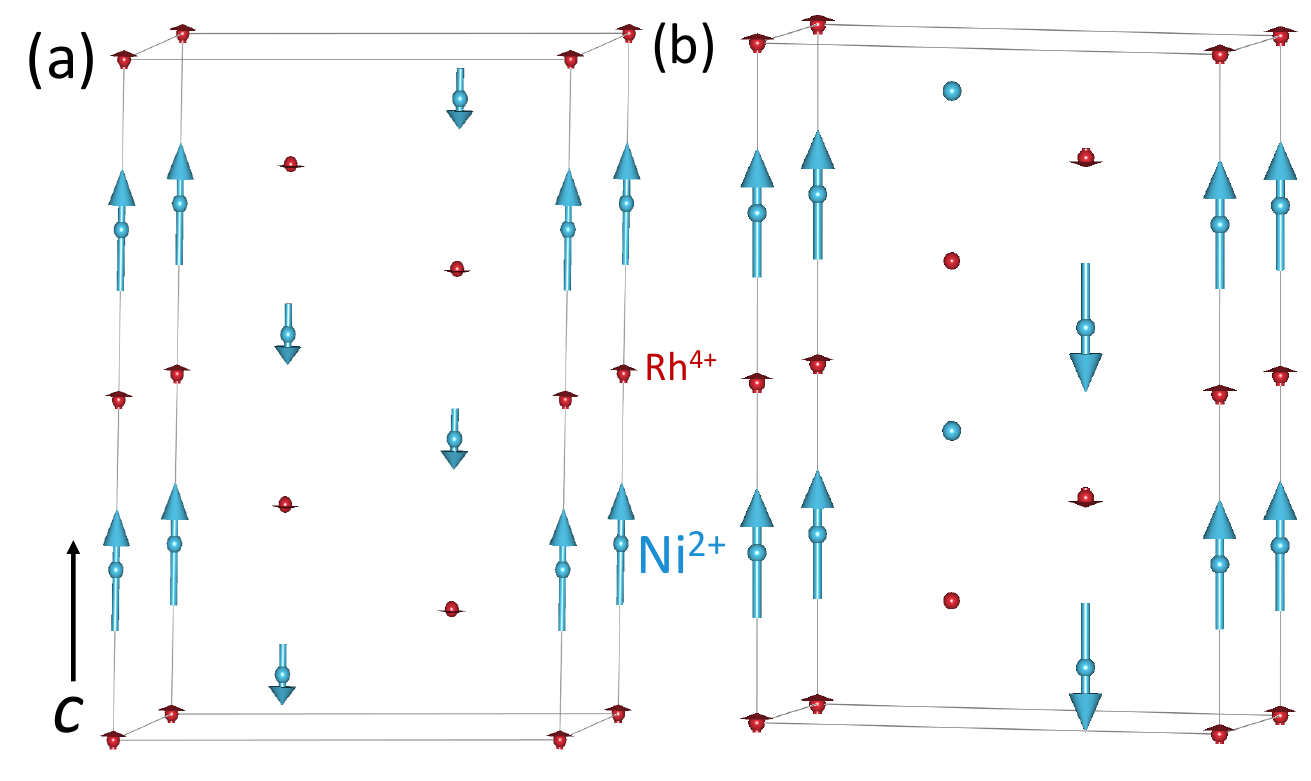}
	\caption{(Color online) Possible magnetic structures of \srni\, below $T_1$ (a) Amplitude modulated  and (b) PDA structure. Chain with no spin in (b) represent incoherent spin-chain with net zero moment.}
	\label{fig:magstr}
\end{figure}

These Fourier coefficients, ${\emph{S}}_{j}^{\textbf{k}}$,  can be written as a linear combination of the basis vectors of the irreducible representation of the propagation vector group \emph{$\textbf{G}$}$_{k}$. 
{In the Bertaut method~\cite{Bertaut1963}, the magnetic structure is derived by coupling the basis vectors corresponding to the same representation at different sites. Therefore, we first consider the representations $\Gamma$$_{1}$ and $\Gamma$$_{2}$ to describe the magnetic structure.
The basis vectors for $\Gamma$$_{1}$ and $\Gamma$$_{2}$  representations are shown in Table~\refeq{basisvector}.  In the $\Gamma$$_{1}$  representation,  spins along the $c$-axis are antiferromagnetically coupled, while in the $\Gamma$$_{2}$  spins along the $c$-axis are ferromagnetically coupled.}
The observed magnetic structure of the present compound can be fitted using  the  representation  $\Gamma_{2}$. In the magnetic structure magnetic moments, in a spin-chain, are aligned ferromagnetically along the crystallographic \emph{c} axis, however, spin-chains are coupled antiferromagnetically in the \textit{ab}-plane with net zero moment in the unit cell. The refined values of ordered magnetic moment for 6$a$ (Ni$^{2+}$) and 6$b$ (Rh$^{4+}$) sites are  0.88 (1) and 0.12 (1) $\mu_B$, respectively.   The observed value of magnetic moment is much smaller than theoretically expected value of 2 and 1 $\mu_B$ for Ni$^{2+}$ and Rh$^{4+}$, respectively (assuming $g$ =2). {The observed small value of the ordered moment could be attributed to enhanced quantum fluctuations associated with geometrical frustration and the low dimensionality of the present system.} The symmetry analysis mandates that an amplitude modulated structure with no vanishing moment at any site (Fig.~\ref{fig:magstr}(a)) differs from a partially disordered antiferromagnetic (PDA) structure  (Fig.~\ref{fig:magstr}(b)) by a global phase that cannot be determined using  powder diffraction. Nevertheless,  equivalence of the two magnetic structure models was validated without relying on symmetry assumptions. Therefore, we also fitted the observed neutron diffraction pattern using the space group $P\bar{1}$, and propagation vector $k = (0, 0, 0)$ and obtained similar chi-square ($\chi^2$) values for the amplitude modulated and PDA configurations. In the PDA state,  2/3 of ferromagnetically coupled spin-chains are ordered in the AFM structure, while the remaining 1/3 are in the incoherent state (with net zero ordered state moment). The schematics of PDA and amplitude modulated structure suitable for \srni\, are shown in Fig.~\ref{fig:magstr}.  
It can be noted that, based on the behavior of ac susceptibility (Appendix~\ref{sec:Acs}) where a large shift in the peak position was observed with frequency, the magnetic ground state can be considered as F-PDA state below 20~K. In the F-PDA state, spins in the 1/3 incoherent chains  of PDA state freeze randomly, indicating that magnetic ground state might not be thermodynamically stable. This F-PDA state is compatible with the PDA structure (Fig.~\ref{fig:magstr}(b))  from neutron diffraction.

\begin{table}[h]
	 {
	\begin{center}
		\caption{\label{basisvector}Basis vectors of positions 6\emph{a} and 6\emph{b} for the representations $\Gamma$$_{1}$ and $\Gamma$$_{2}$.} \vspace{0.3cm}
		\begin{tabular}{lcccc}
			\hline
			\hline
			\\ [-1.5ex]
			IR &\multicolumn{2}{c}{ Basis vectors  for 6\emph{b}  site} &\multicolumn{2}{c}{ Basis vectors  for 6\emph{a}  site}\\
			\\ [-1.5ex]
			&{(0, 0, 0)}&	{(0, 0, 1/2 )}	&{(0, 0, 1/4)}	&{(0, 0, 3/4)} \\
			\hline
			{$\Gamma$$_{1}$} &   {(0 0 1)}    &   {(0 0 -1) }   & {(0 0 1)}    &   {(0 0 -1) }\\
			{$\Gamma$$_{2}$} &   (0 0 1)    &   (0 0 1)    & (0 0 1)    &   (0 0 1) \\
			\\ [-1.5ex]
			\hline
			\hline
	\end{tabular} \end{center}}
\end{table}


\subsection{C. Muon spin relaxation}
\begin{figure}[!htb]
	\centering
	\includegraphics[width = 1.051\linewidth]{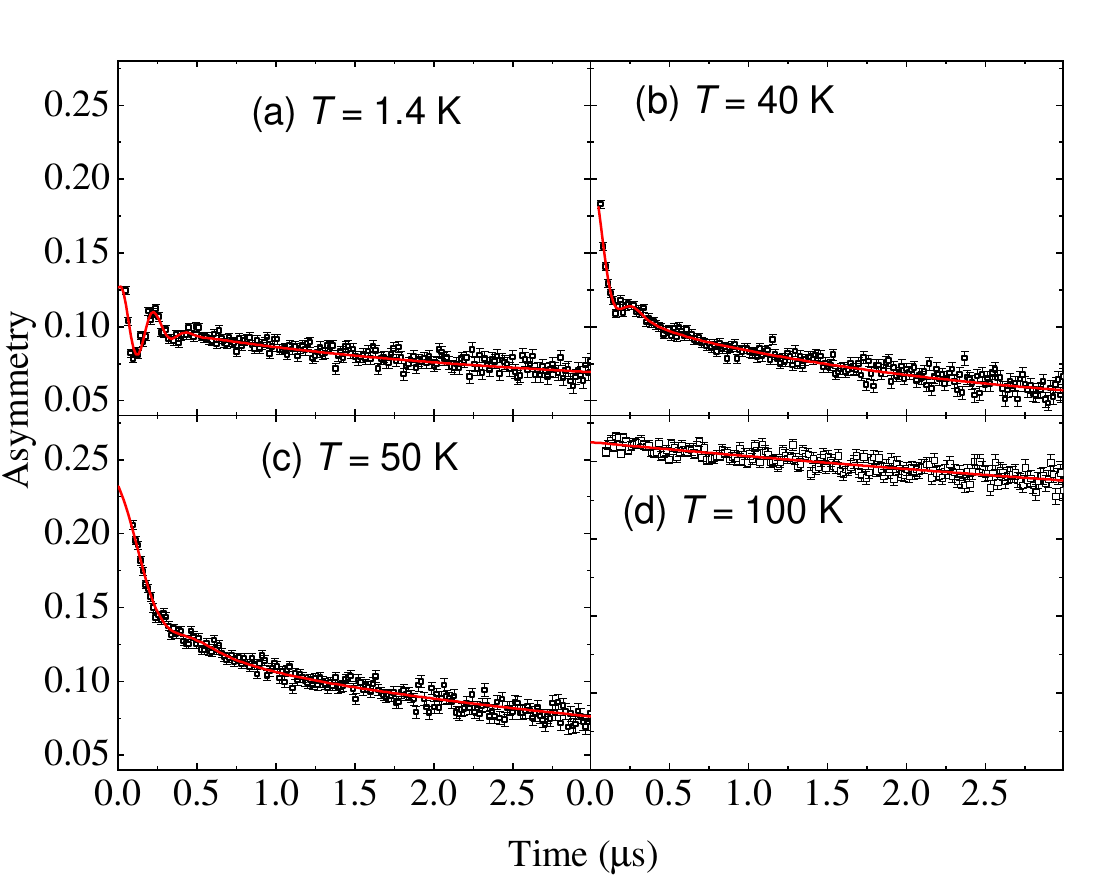}
	\caption{(Color online) (a-d) ZF-$\mu$SR spectrum of \srni{}  at indicated temperatures. The  solid curves are the fits to the $\mu$SR data by Eq.~\ref{muonfitting}.}
	\label{fig:musr}
\end{figure}

\begin{figure}[!htb]
	\centering
	\includegraphics[width = 1.01\linewidth]{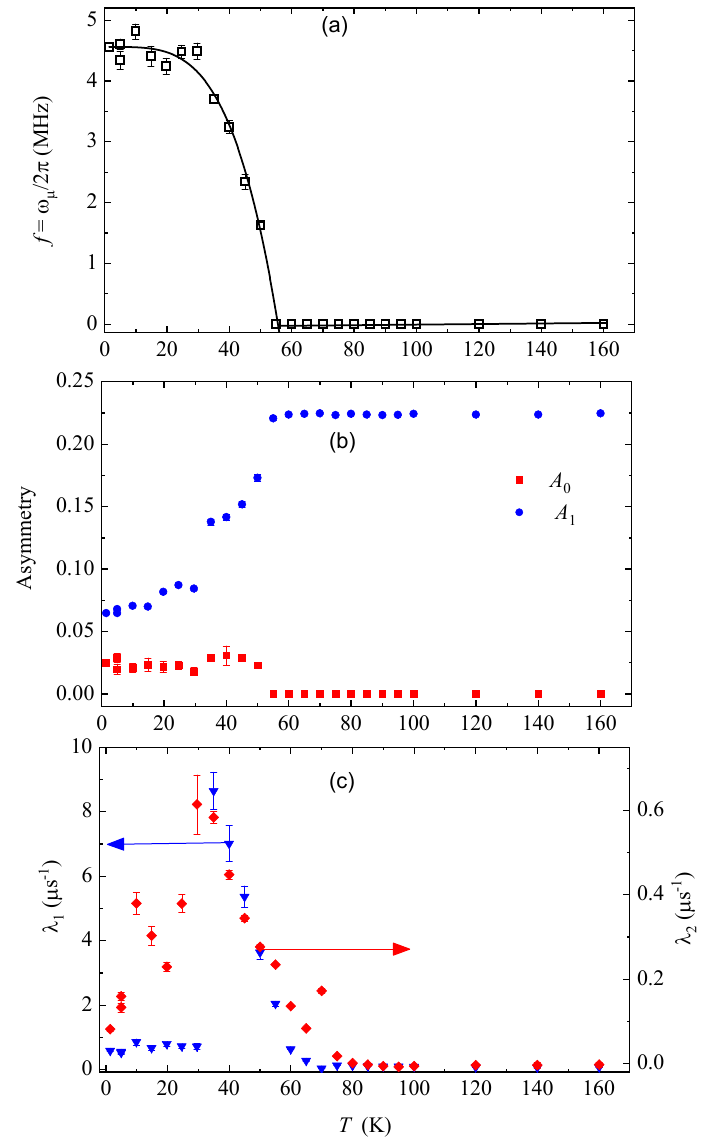}
	\caption{(Color online)  (a) $T$-dependences of muon pre-cession frequency for \srni.  Solid lines are guide to the eye.  (b) $T$-dependences of a normalized ZF asymmetry. (c)  $T$-dependences of characteristic spin lattice relaxation rates $\lambda_{1}$ and $\lambda_{2}$.}
	\label{fig:musr_ass}
\end{figure}

In order to investigate the microscopic nature of magnetic order in \srni\,  we have used the $\mu$SR technique, which is very sensitive to the local magnetic  moment and its environments, because the implanted muons in the sample experience magnetic field of nearest neighbors. Figure~\ref{fig:musr} shows the zero-field (ZF)-$\mu$SR spectrum of \srni\,  at various temperatures over 1.4-100\,K. 	
{
In the paramagnetic regime ($T>T_1$), due to thermal fluctuations, the magnetic moments become disordered, leading to a random orientation of the spins in both time and space. As a result, the expectation value of the local magnetic field at the muon site $<B_\mu>=0$, which prevent coherent muon spin precision. For polycrystalline samples, this results in an exponential relaxation of the muon polarization signal.} Below $T<T_1$,  ZF spectra exhibit two minima (i.e. oscillation with time), indicating the onset of a long-range magnetic ordering. Our analysis shows that there are two muon sites, one site with a field greater than the frequency response of the MuSR spectrometer at ISIS pulsed muon facility due to the pulsed width of 80ns. In the first approximation, the observed spectra seem to be typical for Kubo-Toyabe-type relaxation. However, a careful analysis has shown that a Gaussian damped oscillatory function (mentioned below), probably due to a static order with wide field distribution, is more suitable for fitting the $\mu$SR asymmetry spectra.

\begin{equation}\label{muonfitting}
	\begin{split}
		A_{t} = A_{0} \cos (\omega t + \phi) \rm{exp}(-\sigma^2t^2/2) \\+ A_{1} [w_{1} exp(-\lambda_{1}t) +(1-w_{1})exp(-\lambda_{2}t)] + A_{BG}
	\end{split}
\end{equation}

Here $A_{0}$ is the asymmetry associated with the oscillatory component (some time called transverse component), $A_1$ is the longitudinal component caused by the field component parallel to initial muon spin-polarization and $A_\text{BG}$ is the background contribution to the asymmetry, representing the muons that missed the sample and stopped in the Ag sample holder. The value of  $A_\text{BG}$=0.038 was	estimated from the fitting of 160 K spectra and was kept fixed during the analysis of other temperature data. Further the value of $w$$_{1}$ = 0.5 was kept fixed during the analysis to account the two muon sites, especially in the ordered state. In the paramagnetic state the response from the both the muon stopping sites was assumed identical. Figures~\ref{fig:musr_ass} (a) and (b) show the $T$-dependences of the muon precession frequency $f = \omega_{mu}/2\pi$ and its  asymmetry $A_{0}$, respectively. The characteristic spin lattice relaxation rates $\lambda_{1}$ and $\lambda_{2}$ are shown in Figure~\ref{fig:musr_ass} (c).  Both $f$ and $A_{0}$ have finite value below $\sim$50\,K, around which a kink has been observed in the dc magnetization study  (Fig.~\ref{fig:dcM}(b), in the Appendix~\refeq{sec:appen}) and neutron diffraction (Fig.~\ref{fig:diff}) study shows the presence of long-range magnetic order in PDA state. A change in the spin lattice relaxation rate  $\lambda_{2}$ is clearly visible at $\sim 20~K$, which may be related to the onset of F-PDA state.	A simple dipole field calculation shows several frequency components with a  large standard deviation are possible for the PDA state. This is consistent with the observed Gaussian damped oscillatory function for the $\mu$SR asymmetry spectra.

\subsection{D. Inelastic Neutron Scattering}

\begin{figure}[!htb]
	\centering
	\includegraphics[width = \linewidth]{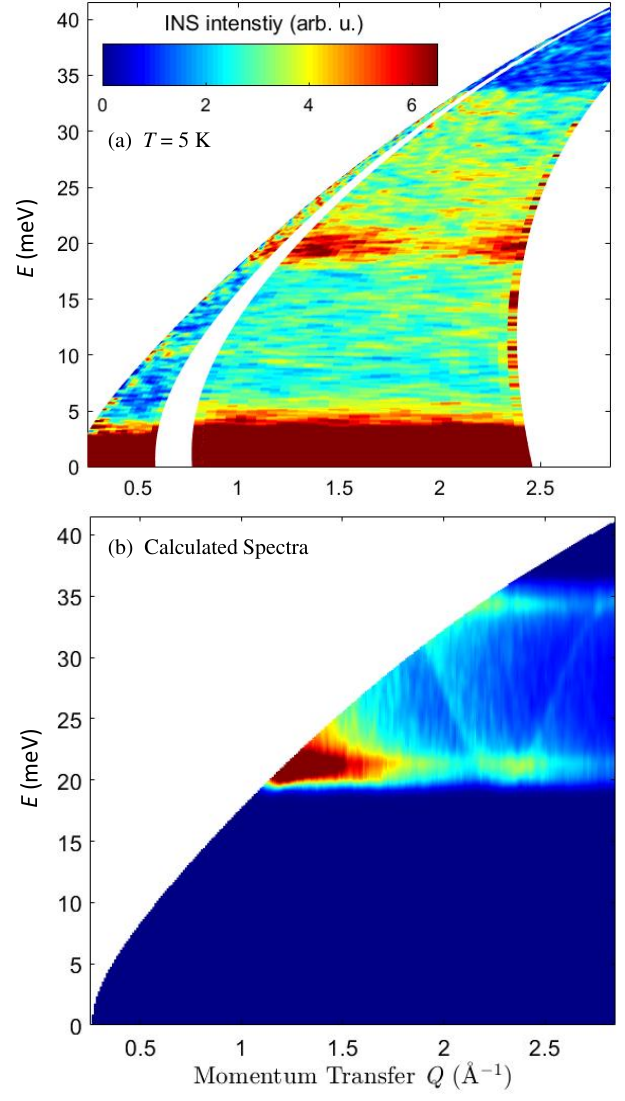}
	\caption{(Color online) (a) Color coded inelastic neutron scattering intensity maps, energy transfer vs momentum transfer ($Q$) of \srni\ measured with an incident energy $E_i=50$ meV on HET. (b) The simulated spin-wave spectra using SpinW program~\cite{Toth_2015} with exchange parameters, $J=3.7$~meV and a single ion anisotropy $D = 10$~meV for Ni$^{2+}$ spins.}
	\label{fig:ins}
\end{figure}

\begin{figure*}
	\includegraphics[height = 0.6\linewidth, width = 0.98\linewidth]{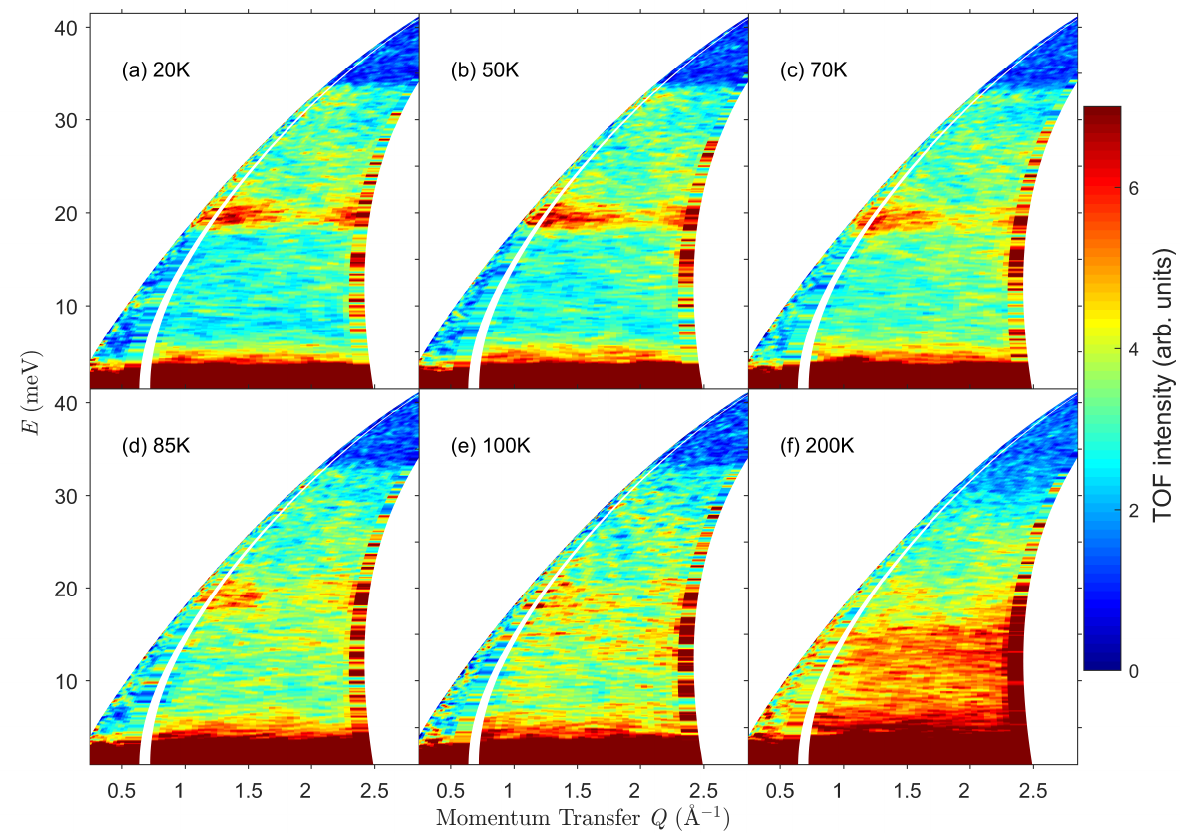}
	\caption{(Color online) (a-f) Temperature evolution of the INS spectra measured over $20 –200$~K.}
	\label{fig:instdep}
\end{figure*}

To estimate the value of magnetic exchange interactions between localized electron magnetic moments and anisotropy parameters, we have performed INS experiments on \srni{}. Figure\,\ref{fig:ins}(a)  shows  the INS response, a color-coded contour map of the intensity, energy transfer
$E$ versus momentum transfer $Q$ for  \srni\ measured at  $T=5$~K (with $E_i=50$ meV) from the low-angle detectors banks up to $Q=3~\AA^{-1}$. The INS spectra show a spin-wave gap of 20~meV and bandwidth of 34\,meV. The intensity of the INS signal, energy integrated over 18-23\,meV, decreases  with increasing  value of $Q$ and follows the form factor of Ni$^{2+}$ ($I\propto |f_{Ni}(Q|^2$)), which is very similar to that observed in Sr$_3$NiIrO$_6$~\cite{Toth2016}. This confirms that the magnetic signal arises from Ni$^{2+}$ spin waves.  To analyze the observed spin wave excitation, we have used the following phenomenological  spin Hamiltonian, containing terms only having intra-chain exchange interactions:
\begin{equation}
	H=-\sum_{i}[J \mathbf{S}_i \cdot \mathbf{S}_{i+1}
	+ DS_{z}^2]
	\label{eq:Hamiltonian}
\end{equation}

Here $i$ is the site index for the Ni$^{2+}$ (TP) ions in the chain, $J$ represents the isotropic Heisenberg coupling between the Ni$^{2+}$ spins, and the $z$ axis is directed along the chains.  $D>0$ is the single-ion anisotropy parameter, and $S_z$ refers to the spin component along the $c$-axis (chain direction). We have not included interchain interactions 
in the phenomenological  spin Hamiltonian because  they are expected to be very weak and thus do not affect the spin-wave dispersion along the chains significantly.
In addition, we have not observed any feature associated with  Rh$^{4+}$ spin-waves and/or  Ni-Rh  intrachain exchange interactions in magnetic excitation spectra, probably due to very small signals and/or these features are present at higher energy as seen for Ir mode near $\sim 90$~meV  in Sr$_3$NiIrO$_6$~\cite{Lefran2016}.	
From the observed spectrum, we can set an upper limit of 0.03 meV for the inter-chain AFM interaction. 	Within the framework of linear spin-wave theory, the dispersion for the spin Hamiltonian~(\ref{eq:Hamiltonian}) is given by :

\begin{equation}
	\hslash \omega_L = 2S [(D + J) - J \cos({\pi}L) ]\,.
	\label{Eq:Dispersion}
\end{equation}

Figure~\ref{fig:ins}(b) shows the powder spectrum simulated by linear spin-wave theory for $J = 3.7$~ meV and $D=10$\,meV using spinW~\cite{Toth_2015}, which reveals good agreemnt with the observed spin-wave excitations. The temperature dependence of INS spectra is shown in Fig.~\ref{fig:instdep}. Surprisingly, the spin wave excitation survive up to $T=85$~K (1.5 times $T_{1}$), suggesting that intrachain FM correlation builds much above magnetic ordering temperature. Similar behavior has been observed for other compound of  $A_{3}MX$O$_{6}$ family such as  Ca$_3$Co$_2$O$_6$ \cite{Jain2013,Agrestini2014} and {Sr$_3$NiIrO$_6$}~\cite{Toth2016} (spin-1 and spin-1/2 alternating chain).  The observed behavior can be attributed to a quasi-1D nature of the magnetic interactions in \srni. Upon heating above 85~K, we observe a reduction of the gap and a transfer of spectral weight from higher energy  into the spin-gap region.

{\subsection{E. Domain Wall Formation and Dynamics}}

Having determined the value of exchange and anisotropy parameters, we now discuss the low energy magnetic excitation in \srni.  From the observed large value of $D$ ($D>>J$), it is clear that  the lowest energy excitation in \srni, should be a single domain-wall (assuming negligible interchain exchange interaction $J_\text{inter}$) with energy cost $2JS^2$ (85~K), which can move easily along the chain without any energy cost. The single domain-wall here refers to the two FM regions of opposite polarity. At finite temperature,  the total free energy of the system has contribution from entropy ($ S = k_{B}T\,log_{e} N $)  also the resulting free energy, therefore,  is $F = 2JS^{2} - k_{B}T\,log_{e} N$.  An important point to be noted here is that for FM chains, the domain wall can propagate only via a thermally  activated process. {These thermally activated domain walls lead to high degeneracy in the PDA state at finite temperatures, resulting in more domain wall formation. However, at low temperatures, their propagation becomes extremely slow.} Due to this, fluctuating high-temperature multidomains  may  lead to randomness and disorder  at lower temperatures. This  results in a transition from PDA to F-PDA-like state, as observed for Ca$_{3}$CoRhO$_{6}$~\cite{Sampathkumaran2002,Niitak177202}. The  observed frequency dependence of real $\chi_{ac}^{'}$  and imaginary $\chi_{ac}^{''}$  components of ac susceptibility (Appendix~\refeq{sec:appen}, Figs.~\ref{fig:chiacfc} and \ref{fig:chiaczf}) and its persistence under an applied dc magnetic field of $\sim70$\,kOe, with only a small reduction in its intensity,  at $\sim20~$K, could be consistently interpreted with
the idea of F-PDA-like state.

\section{CONCLUSIONS}	

In conclusion, we have investigated the magnetic ground state in \srni{{} using muon spin rotation and relaxation ($\mu$SR), neutron diffraction and inelastic neutron scattering techniques. {The neutron diffraction study, combined with ac susceptibility and $\mu$SR measurements, indicates the presence of a partially-disordered antiferromagnetic (PDA) state below $T_1$.} Below $T_{2}$, the 1/3 incoherent  spin-chains of PDA structure freeze into a F-PDA-like state (Appendix~\refeq{sec:Acs}).  Our INS study showed  dispersing magnetic excitations, whose quasi-1D character evidences	the prevalence of FM intrachain (anisotropy gap $D$ three times larger than the exchange interation $J$) over AFM interchain interactions.  The observed large value of $D$ ($D>>J$) in these triangular lattice antiferromagnets, with FM intrachain interaction,  results in the formation of domain walls.  Freezing of domain walls results into formation  of  a F-PDA-like state below  $T_{2}$. The present study shows that a F-PDA-like state, as the ground state of the triangular lattice antiferromagnetic system,  presents a complicated spin configuration due to geometrical frustration at low temperatures. The values of exchange parameters obtained in the  present study will be helpful to carry out large scale Monte Carlo calculations to estimate the equal time correlation functions and for the development of theoretical models to understand  the domain wall dynamics of the PDA state in geometrically frustrated magnets.

	\section*{acknowledgments}	
	D.T.A. thanks the EPSRC UK
	for funding (Grant No. EP/W00562X/1), the Royal Society of London for International
	Exchange funding between the UK and Japan, and Newton
	Advanced Fellowship funding between the UK and China and the CAS for PIFI Fellowship. SMY acknowledges the financial assistance from SERB, Department of Science and Technology, Government of India under the J C Bose fellowship program (JCB/2023/000014).

	%
	%
	\appendix	 	
	\label{sec:appendix}
	\section{Dc Magnetization}
	\label{sec:appen}
	\begin{figure}[]
		\centering
		\includegraphics[width = 0.9\linewidth]{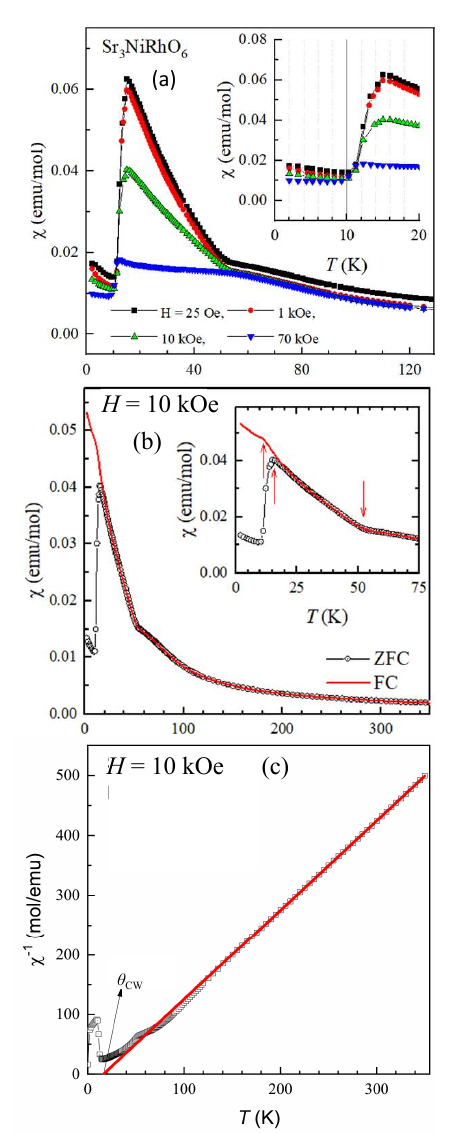}
		\caption{(Color online) (a) Temperature variation of the dc magnetization under various applied magnetic field for \srni{}. (b) Temperature dependence of the field cooled (FC) and zero field cooled (ZFC) magnetization for \srni{} under applied magnetic field of 10~kOe. (c) Temperature variation of high temperature inverse magnetic susceptibility under applied magnetic field of 10~kOe.}
		\label{fig:dcM}
	\end{figure}
	
	\begin{figure*}[]
		\includegraphics[height = 1.09\linewidth, width = 0.9\linewidth]{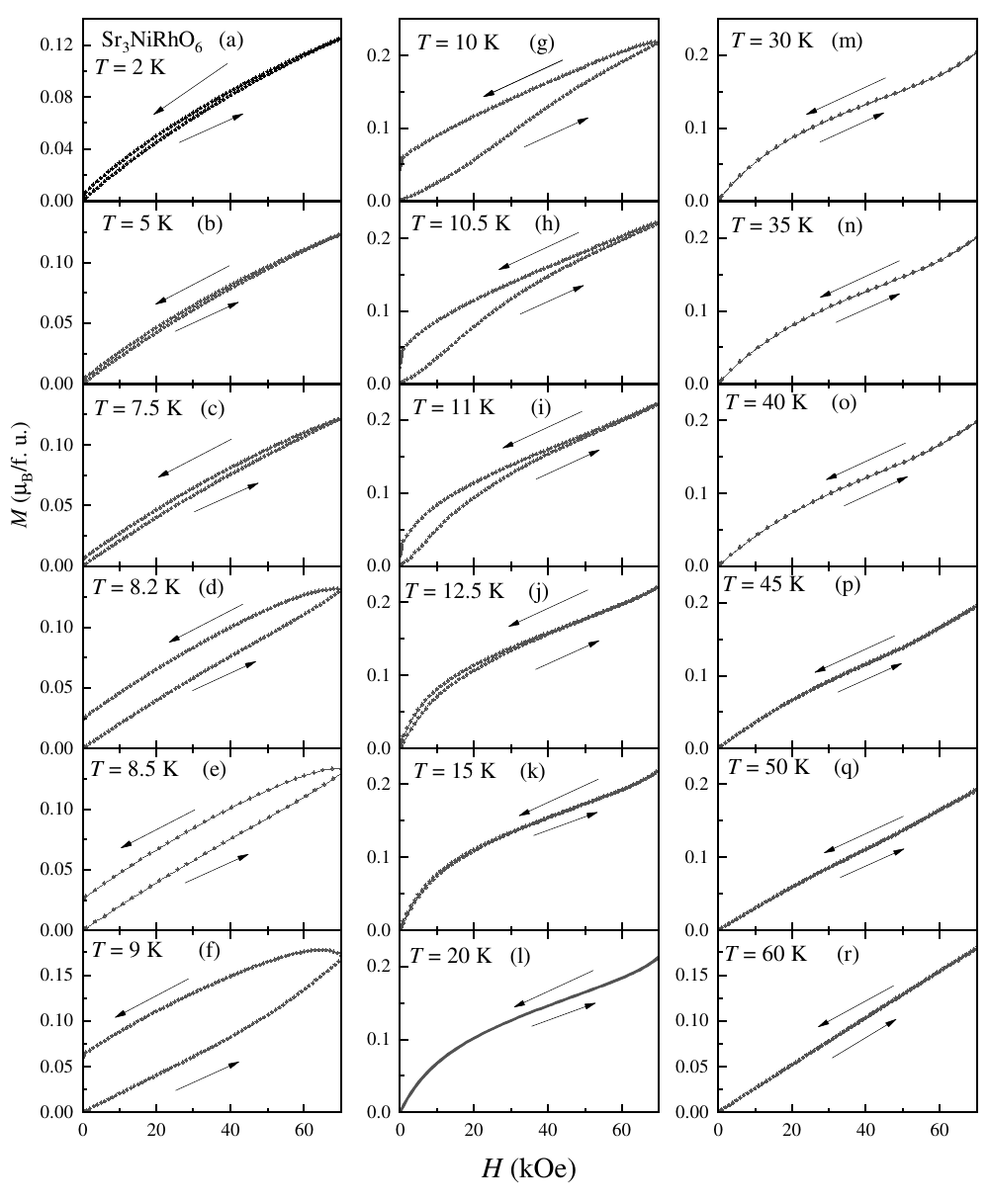}
		\caption{(Color online) Isothermal  dc magnetization for \srni{} at various temperatures. The arrows indicate direction of external magnetic field variation.}
		\label{fig:mh}
	\end{figure*}

	\begin{figure}[]
		\centering
		\includegraphics[width = 1\linewidth]{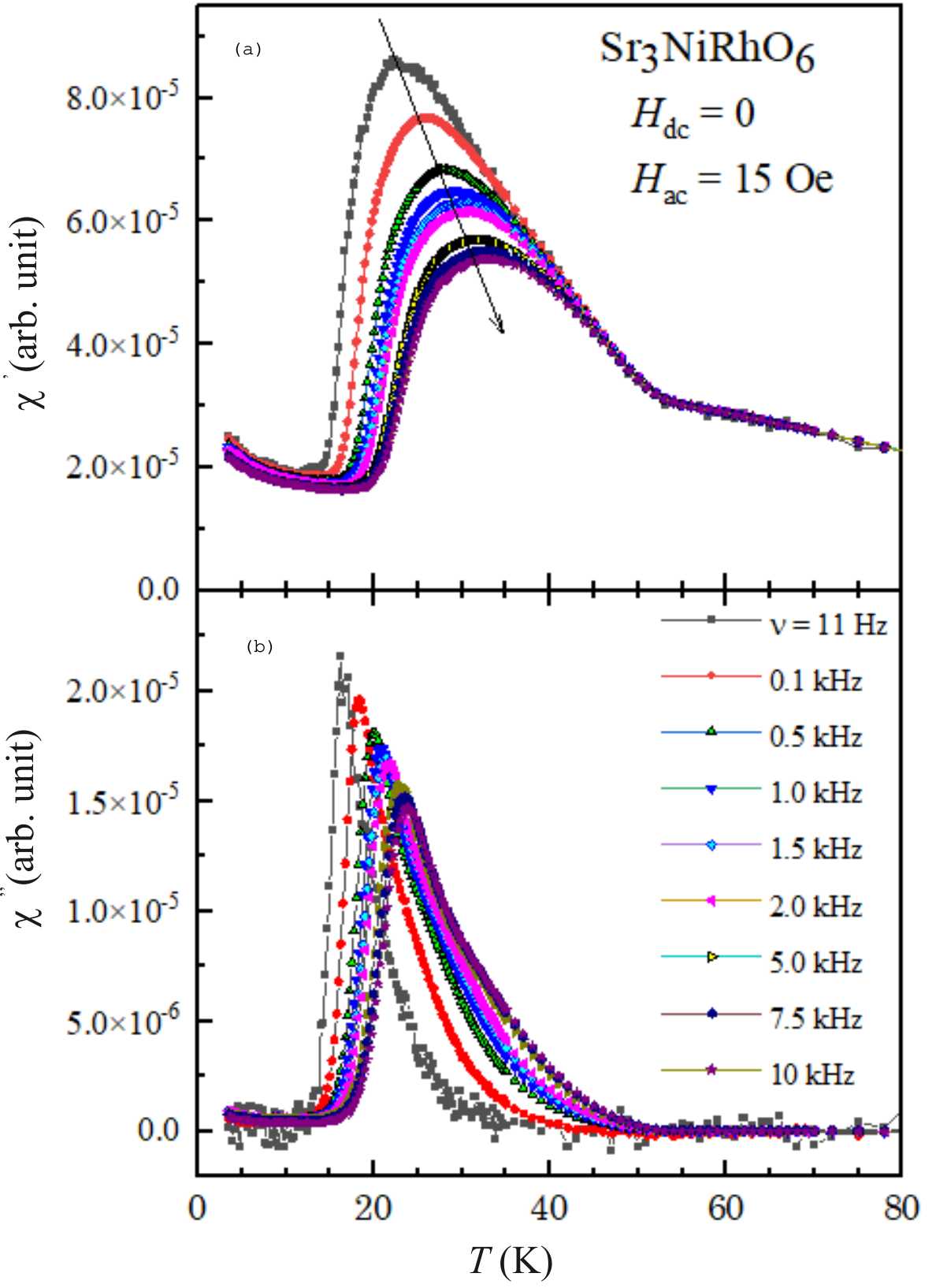}
		\put(-110,290){{\bf(a)}}
		\put(-110,140){\bf(b)}
		\caption{(Color online) The temperature dependence of (a) real and (b) imaginary parts of ac susceptibility for \srni{} at varied frequencies under zero applied dc magnetic field.}
		\label{fig:chiaczf}
	\end{figure}

	\begin{figure}[]
		\centering
		\includegraphics[width = 1\linewidth]{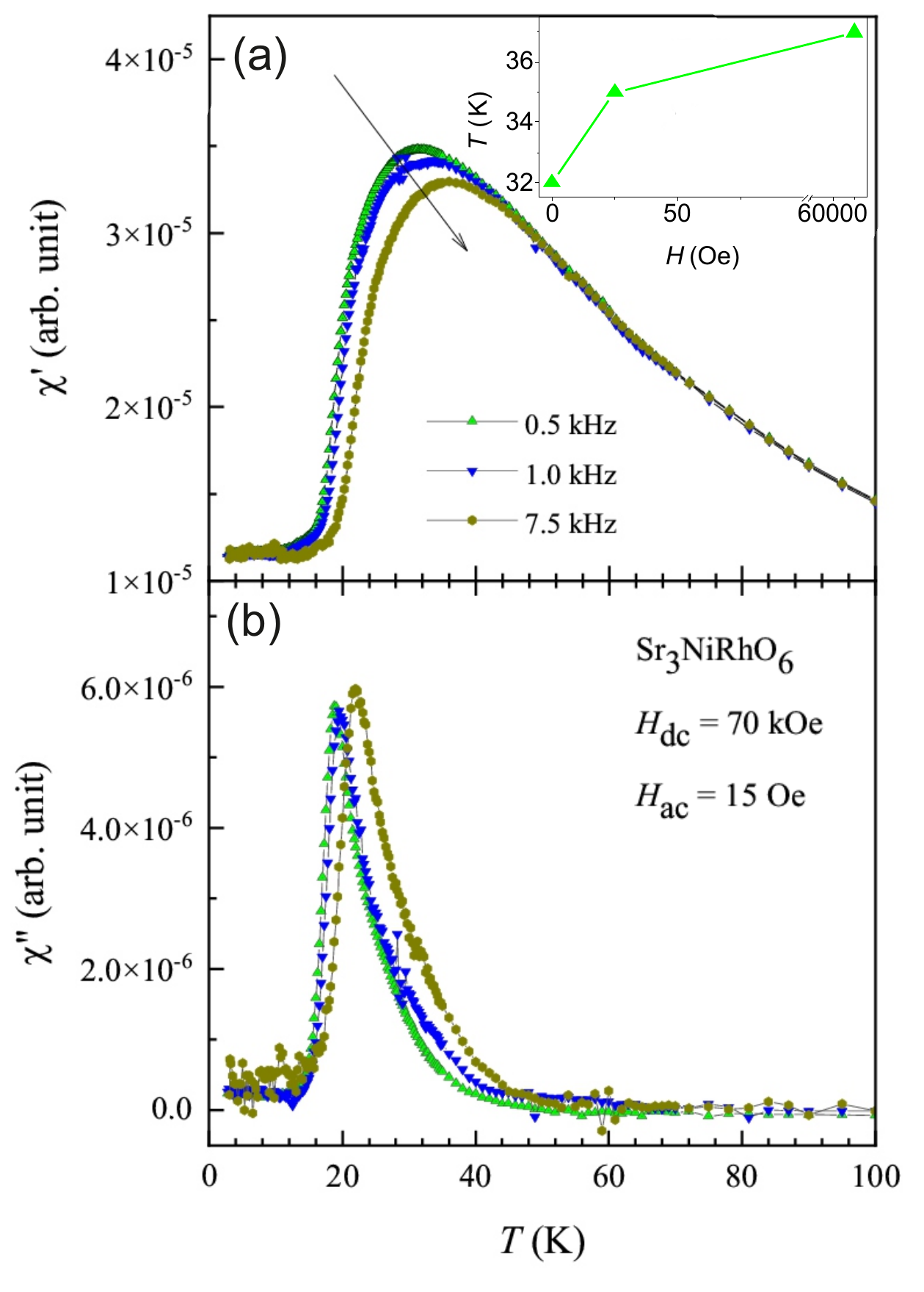}
		\caption{(Color online)  The temperature dependence of (a) real and (b) imaginary parts of ac susceptibility for \srni{} at varied frequencies under dc magnetic field of 70~kOe. Inset in (a) shows field variations of the center of the  peak observed in $\chi_{ac}^{'}$.}
		\label{fig:chiacfc}
	\end{figure}

	Figure~\ref{fig:dcM} shows the dc magnetization measurements for \srni{} under various applied magnetic field of ~1, 25, 10, and 70~kOe. Under zero field cooled (ZFC) condition [Fig~\ref{fig:dcM}(a))],  magnetization increases gradually with decreasing the temperature. A kink is observed in the ZFC magnetization at $\sim$~50~K ($T_{1}$), indicating the onset of a magnetic phase transition [(Fig~\ref{fig:dcM}(b))]. 	Both field cooled (FC) and ZFC magnetizations increase sharply below $T_{1}$.  At low temperature $\sim$15~K($T_{2}$), a sharp drop in the ZFC magnetization along with a bifurcation in the FC and ZFC magnetization has been observed, which could be due to (i) a possible spin freezing due to the presence of competing exchange interactions and (ii) domain wall freezing.   In addition, at $\sim$10~K, a small kink has also been observed in the magnetization curves. 
	Similar two characteristic temperatures $T_{1}$ and $T_{2}$ have been reported in literature for the other compounds of  the $A_{3}MXO_{6}$ family, such as~Ca$_\text{3}$CoRhO$_\text{6}$~\cite{Niitak177202,Niitaka2001, Loewenhaupt2003},  and Sr$_\text{3}$NiIrO$_\text{6}$~\cite{Flahaut2003,Mikhailova2012}. For both compounds, neutron diffraction studies have confirmed a partially disordered antiferromagnetic (PDA) state or amplitude modulated structure [in latter case~\cite{Lefrancois2014})].
	
	High temperature inverse magnetic susceptibility ($\chi^{-1}$), shown in Fig~\ref{fig:dcM}(c), follows Curie-Weiss law with a paramagnetic Curie temperature ($\theta_\text{CW}$) of 16.2~K. The derived value of the effective moment from the slope of the fitting results in 2.3~$\mu_{\rm B}$/f.u. The observed value of the effective paramagnetic moment is less than the theoretically expected  value of 3.2~$\mu_{\rm B}$/f.u, considering $S=~1/2$ and 1 for Rh$^{4+}$ and Ni$^{2+}$, respectively (with $g$ = 2). The observed small value of the magnetic moment could be due to strong covalent mixing, expected in 4$d$ transition metal oxides. Since $\theta_\text{CW}$ represent the combined effect of all exchange interactions, the observed positive value  indicates that dominant exchange interaction in \srni~ is FM.

	To understand the magnetization behavior  in \srni{}, we have measured dc magnetization as function of applied magnetic field at various temperatures over 1.5-60~K (Fig.~\ref{fig:mh}). A paramagnetic-like behavior has been observed above $T_\text{1}$ (50~K). Below $T_\text{1}$, a curvature has been observed in the magnetization curves, which becomes prominent below 40~K. For $10.5 \le T \le 20~\text{K}$, a step like shape of the magnetization is clearly visible.  The value of the magnetization in the plateau region is $\sim0.2\,{\mu}_\text{B}$, which is close to 1/3 of the expected value of the full ordered moment (obtained from neutron diffraction study discussed in the main text) considering \textcolor{red} {a} fully polarized FM state, with parallel alignment  of all Rh$^{4+}$ and Ni$^{2+}$ spins in a unit cell. The observed step-like behavior could be due to a magnetic field induced transition to a ferrimagnetic state, where two thirds of the spin-chains have spin up and the remaining one third have spin down. For $8.25 \le T \le11~\text{K}$, the revised leg of the $M(H)$ curve, while approaching a zero field, shows that hysteresis  is present. The observed hysteresis  occupies all field range for $8.2 < {T}<10.5~\text{K}$, however, it is very weak for 
	$2 \le T \le 7.5 ~\text{K}$,  with no saturation of the magnetization under highest applied magnetic field of 7~T. 
	
	\section{Ac Susceptibility}
	\label{sec:Acs}
	
	To probe the origin of the bifurcation in the  FC and ZFC magnetization and get a better insight of the magnetic dynamics in \srni{}, we have measured ac susceptibility as a function of temperature at various frequencies. A peak in the ac susceptibility has been observed (Fig.~\ref{fig:chiaczf}) at $\sim$20~K in both real and  imaginary part of the ac susceptibility.  Peak shifts to the higher temperature  by increasing the frequency of the applied ac field indicated the onset of glassiness around 20~K. Surprisingly, the observed shift in the position of the ac susceptibility peak is much larger than expected for a conventional spin-glass system.  It may be remarked that under applied dc magnetic field of 70~kOe,  the peak in both $\chi_{ac}^{'}$ and  $\chi_{ac}^{''}$ persists, with only a small reduction in its intensity. These observations reveal that the  $\chi_{ac}^{'}$ signal in zero field is made up of a FM part and a spin-glass part. Under applied dc magnetic field of 70~kOe, the spin-glass part gets suppressed, however, the FM part behave like clusters (created by negligible motion of domain wall)  with some glassy dynamics.

	{
	\section{Additional Neutron Diffraction Patterns}}
	\label{sec:append3}
	 
	 A neutron diffraction pattern has been recorded over wide $Q$-range (Fig.~\ref{Fig13:GEM}) using  GEM TOF diffractometer at ISIS Facility, to examine if there is any anti-site mixing for Ni/Rh. The analysis has revealed 
  no improvement of the fit when varying the occupancies of both Ni and Rh from their expected values.  This rules out the possibility of anti-site mixing in our sample.  
	
\begin{figure}[h]
	\centering
	\includegraphics[width = 1\linewidth]{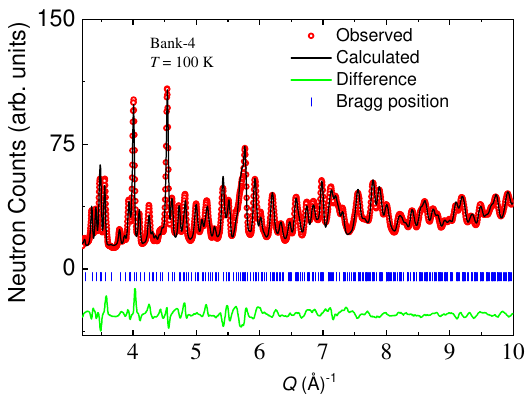}
	\caption{{(Color online) Observed (open circles) and calculated (solid lines) neutron diffraction patterns at 100~K for \srni{} over high values of $Q$ from the GEM TOF diffractometer. The green solid line at the bottom shows the difference between the observed and the calculated patterns. Vertical lines show the positions of the nuclear Bragg peaks.}}
	\label{Fig13:GEM}
\end{figure}

\clearpage
	\bibliography{library}
	
\end{document}